\documentclass[12pt]{article}
\usepackage{epsfig}
\usepackage{xspace}
\usepackage{hhline}
\usepackage{amssymb}
\usepackage{times}
\usepackage{cite}

\newlength{\dinwidth}
\newlength{\dinmargin}
\begin{document}  
\newcommand{\pom}{{I\!\!P}}
\newcommand{\reg}{{I\!\!R}}
\newcommand{\slowpi}{\pi_{\mathit{slow}}}
\newcommand{\fiidiii}{F_2^{D(3)}}
\newcommand{\fiidiiiarg}{\fiidiii\,(\beta,\,Q^2,\,x)}
\newcommand{\n}{1.19\pm 0.06 (stat.) \pm0.07 (syst.)}
\newcommand{\nz}{1.30\pm 0.08 (stat.)^{+0.08}_{-0.14} (syst.)}
\newcommand{\fiidiiiful}{F_2^{D(4)}\,(\beta,\,Q^2,\,x,\,t)}
\newcommand{\fiipom}{\tilde F_2^D}
\newcommand{\ALPHA}{1.10\pm0.03 (stat.) \pm0.04 (syst.)}
\newcommand{\ALPHAZ}{1.15\pm0.04 (stat.)^{+0.04}_{-0.07} (syst.)}
\newcommand{\fiipomarg}{\fiipom\,(\beta,\,Q^2)}
\newcommand{\pomflux}{f_{\pom / p}}
\newcommand{\nxpom}{1.19\pm 0.06 (stat.) \pm0.07 (syst.)}
\newcommand {\gapprox}
   {\raisebox{-0.7ex}{$\stackrel {\textstyle>}{\sim}$}}
\newcommand {\lapprox}
   {\raisebox{-0.7ex}{$\stackrel {\textstyle<}{\sim}$}}
\def\gsim{\,\lower.25ex\hbox{$\scriptstyle\sim$}\kern-1.30ex%
\raise 0.55ex\hbox{$\scriptstyle >$}\,}
\def\lsim{\,\lower.25ex\hbox{$\scriptstyle\sim$}\kern-1.30ex%
\raise 0.55ex\hbox{$\scriptstyle <$}\,}
\newcommand{\pomfluxarg}{f_{\pom / p}\,(x_\pom)}
\newcommand{\dsf}{\mbox{$F_2^{D(3)}$}}
\newcommand{\dsfva}{\mbox{$F_2^{D(3)}(\beta,Q^2,x_{I\!\!P})$}}
\newcommand{\dsfvb}{\mbox{$F_2^{D(3)}(\beta,Q^2,x)$}}
\newcommand{\dsfpom}{$F_2^{I\!\!P}$}
\newcommand{\gap}{\stackrel{>}{\sim}}
\newcommand{\lap}{\stackrel{<}{\sim}}
\newcommand{\fem}{$F_2^{em}$}
\newcommand{\tsnmp}{$\tilde{\sigma}_{NC}(e^{\mp})$}
\newcommand{\tsnm}{$\tilde{\sigma}_{NC}(e^-)$}
\newcommand{\tsnp}{$\tilde{\sigma}_{NC}(e^+)$}
\newcommand{\st}{$\star$}
\newcommand{\sst}{$\star \star$}
\newcommand{\ssst}{$\star \star \star$}
\newcommand{\sssst}{$\star \star \star \star$}
\newcommand{\tw}{\theta_W}
\newcommand{\sw}{\sin{\theta_W}}
\newcommand{\cw}{\cos{\theta_W}}
\newcommand{\sww}{\sin^2{\theta_W}}
\newcommand{\cww}{\cos^2{\theta_W}}
\newcommand{\trm}{m_{\perp}}
\newcommand{\trp}{p_{\perp}}
\newcommand{\trmm}{m_{\perp}^2}
\newcommand{\trpp}{p_{\perp}^2}
\newcommand{\alp}{\alpha_s}

\newcommand{\alps}{\alpha_s}
\newcommand{\sqrts}{$\sqrt{s}$}
\newcommand{\LO}{$O(\alpha_s^0)$}
\newcommand{\Oa}{$O(\alpha_s)$}
\newcommand{\Oaa}{$O(\alpha_s^2)$}
\newcommand{\PT}{p_{\perp}}
\newcommand{\JPSI}{J/\psi}
\newcommand{\sh}{\hat{s}}
\newcommand{\uh}{\hat{u}}
\newcommand{\MP}{m_{J/\psi}}
\newcommand{\PO}{I\!\!P}
\newcommand{\xbj}{x}
\newcommand{\xpom}{x_{\PO}}
\newcommand{\ttbs}{\char'134}
\newcommand{\xpomlo}{3\times10^{-4}}  
\newcommand{\xpomup}{0.05}  
\newcommand{\dgr}{^\circ}
\newcommand{\pbarnt}{\,\mbox{{\rm pb$^{-1}$}}}
\newcommand{\WBoson}{\mbox{$W$}}
\newcommand{\fbarn}{\,\mbox{{\rm fb}}}
\newcommand{\fbarnt}{\,\mbox{{\rm fb$^{-1}$}}}
%
%
\newcommand{\qsq}{\ensuremath{Q^2} }
\newcommand{\gev}{\ensuremath{\mathrm{GeV}}\xspace }
\newcommand{\gevsq}{\ensuremath{\mathrm{GeV}^2}\xspace }
\newcommand{\et}{\ensuremath{E_t^*} }
\newcommand{\rap}{\ensuremath{\eta^*} }
\newcommand{\gp}{\ensuremath{\gamma^*}p }
\newcommand{\dsiget}{\ensuremath{{\rm d}\sigma_{ep}/{\rm d}E_t^*} }
\newcommand{\dsigrap}{\ensuremath{{\rm d}\sigma_{ep}/{\rm d}\eta^*} }
\newcommand{\Odd}{\mathbb{O}}
\newcommand{\odd}{{\hspace{0.1em}\textit{\textsf{I}}\hspace{-0.38em} O}}
\newcommand{\oddsmall}{{\hspace{0.05em}I\hspace{-0.43em} O}}
\newcommand{\TC}[2]{\textcolor{#1}{#2}}
\newcommand{\black}[1]{\textcolor{black}{#1}}
\newcommand{\red}[1]{\textcolor{red}{#1}}
\newcommand{\blue}[1]{\textcolor{blue}{#1}}
\newcommand{\green}[1]{\textcolor{green}{#1}}
\newcommand{\magenta}[1]{\textcolor{magenta}{#1}}
\newcommand{\cyan}[1]{\textcolor{cyan}{#1}}
\newcommand{\purple}[1]{\textcolor{purple}{#1}}
\newcommand{\forrest}[1]{\textcolor{forrestgreen}{#1}}
\newcommand{\lowmagenta}[1]{\textcolor{lowmagenta}{#1}}
\newcommand{\lightblue}[1]{\textcolor{lightblue}{#1}}
\def\Journal#1#2#3#4{{#1} {\bf #2} (#3) #4}
\def\NCA{\em Nuovo Cimento}
\def\NIM{\em Nucl. Instrum. Methods}
\def\NIMA{{\em Nucl. Instrum. Methods} {\bf A}}
\def\NPB{{\em Nucl. Phys.}   {\bf B}}
\def\PLB{{\em Phys. Lett.}   {\bf B}}
\def\PRL{\em Phys. Rev. Lett.}
\def\PRD{{\em Phys. Rev.}    {\bf D}}
\def\ZPC{{\em Z. Phys.}      {\bf C}}
\def\EJC{{\em Eur. Phys. J.} {\bf C}}
\def\CPC{\em Comp. Phys. Commun.}

\begin{titlepage}

\noindent
DESY 02-087  \hfill  ISSN 0418-9833 \\
June 2002

\vspace*{3cm}

\begin{center}
\begin{Large}

{\bf Search for Odderon-Induced\\  
    Contributions to Exclusive {\boldmath $\pi^{\circ}$}\\ 
    Photoproduction at HERA\\}

\vspace*{1cm}

H1 Collaboration

\end{Large}
\end{center}

\vspace{3cm}

\begin{abstract}
A search for contributions to the reaction $ep\to e \pi^{\circ}
N^{\ast}$ from photon-Odderon fusion in the photoproduction regime at
HERA is reported, at an average photon-proton centre-of-mass energy
$\langle W\rangle = 215$ GeV.  The measurement proceeds via detection
of the $\pi^{\circ}$ decay photons, a leading neutron from the $N^*$
decay, and the scattered electron. No $\pi^{\circ}$ signal is observed
and an upper limit on the cross section for the photon-Odderon fusion
process of $\sigma(\gamma p\to\pi^{\circ}N^{\ast}) < 49$~nb at the 95~\%
confidence level is derived, 
integrated over the experimentally accessible range
of the squared four-momentum transfer at the nucleon vertex $0.02\ <
|t| < 0.3~\;\gevsq$. This excludes a recent prediction
from a calculation based on
a non-perturbative QCD model
of a photon-Odderon fusion
cross section above 200~nb.

\end{abstract}

\vspace{1.5cm}

\begin{center}
To be submitted to Physics Letters B
\end{center}

\end{titlepage}

\begin{flushleft}

C.~Adloff$^{33}$,              
V.~Andreev$^{24}$,             
B.~Andrieu$^{27}$,             
T.~Anthonis$^{4}$,             
V.~Arkadov$^{35}$,             
A.~Astvatsatourov$^{35}$,      
A.~Babaev$^{23}$,              
J.~B\"ahr$^{35}$,              
P.~Baranov$^{24}$,             
E.~Barrelet$^{28}$
% ...

\bigskip{\it
% ...
$ ^{4}$ Inter-University Institute for High Energies ULB-VUB, Brussels;
          Universiteit Antwerpen (UIA), Antwerpen; Belgium$^{ c}$ \\
% ...
$ ^{23}$ Institute for Theoretical and Experimental Physics,
          Moscow, Russia$^{ l}$ \\
 $ ^{24}$ Lebedev Physical Institute, Moscow, Russia$^{ e}$ \\
% ...
$ ^{27}$ LPNHE, Ecole Polytechnique, IN2P3-CNRS, Palaiseau, France \\
 $ ^{28}$ LPNHE, Universit\'{e}s Paris VI and VII, IN2P3-CNRS,
          Paris, France \\
% ...
$ ^{33}$ Fachbereich Physik, Bergische Universit\"at Gesamthochschule
          Wuppertal, Wuppertal, Germany \\
% ...
$ ^{35}$ DESY, Zeuthen, Germany \\
% ...
$ ^c$ Supported by FNRS-FWO-Vlaanderen, IISN-IIKW and IWT \\
% ...
$ ^e$ Supported by the Deutsche Forschungsgemeinschaft \\
% ...
$ ^l$ Partially Supported by Russian Foundation
      for Basic Research, grant    no. 00-15-96584 \\
}

\end{flushleft}

\newpage

\section{Introduction}
\noindent
Despite the many successes of quantum-chromodynamics (QCD) in describing
hard strong interactions, the bulk of hadronic cross sections remain
relatively poorly understood.  
A conjecture by Pomeranchuk, known as the Pomeranchuk
theorem~\cite{pom}, states that, for asymptotically large energies,
the difference between hadron-hadron and hadron-antihadron total cross
sections vanishes.  This behaviour is explained by the dominant
exchange of the Pomeranchuk trajectory, the ``Pomeron'' $\pom$,
between the scattering particles. The Pomeron trajectory carries the
quantum numbers of the vacuum and is characterized by an intercept
$\alpha_{\pom}(0)\approx 1.08$~\cite{Donnachie:1992ny}, leading to an
approximate energy independence of the elastic and - via the optical
theorem - the total cross sections ($\sigma_{\rm tot} \sim
s^{\alpha_{\pom}(0)-1}$, $s$ being the square of the centre of mass 
energy).  It has been suggested, however, that a
partner of the Pomeron with odd parity $P$ and charge conjugation
parity $C$, the ``Odderon''
$\odd$~\cite{Lukaszuk:1973nt,Joynson:1975az,Kang:1974gt}, exists.
Such an additional $C=P=-1$ exchange contributes with
opposite signs to the particle-particle and particle-antiparticle
scattering amplitudes, creating a finite cross section difference at
high energy if the corresponding Odderon trajectory has an intercept
$\alpha_{\oddsmall}(0)$ close to 1.  However, in the explored energy
range and within the accuracy of the present data~\cite{pdg2000}, no
difference remains at high energies between the measured total cross
sections for proton-proton and proton-antiproton interactions. Hence,
any difference between the cross sections must be small, necessitating a
more sensitive search for the Odderon.  Within QCD,
the Pomeron is modelled, to lowest order, as a two gluon
exchange in a net colour singlet state. Similarly a 
net colour singlet three gluon exchange,
which is predicted by QCD, can be associated with the Odderon.  In
perturbative QCD exact solutions for the Odderon intercept have been
found~\cite{Janik:Bartels}.  The search for the Odderon has therefore
become an additional part of the QCD tests to be performed at HERA,
and expectations for its discovery are high.

\begin{figure}[htp]
  \center
  \epsfig{file=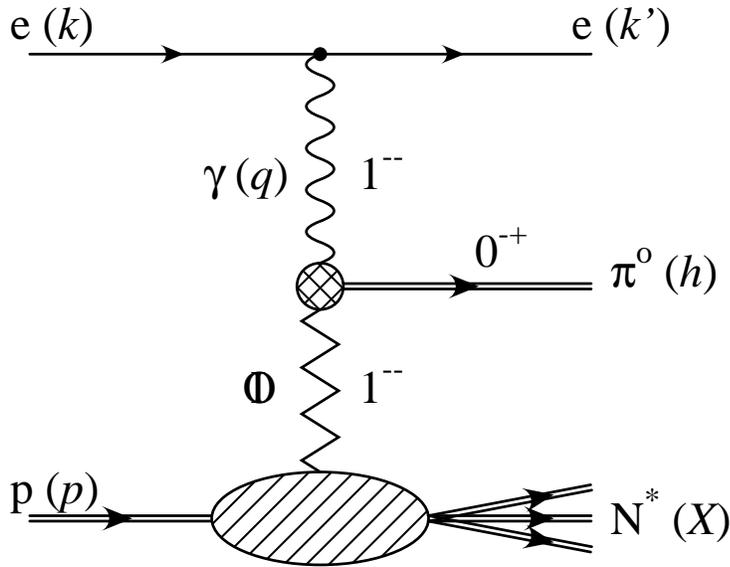,width=11cm}\\
  \caption{Diagram for the process $ep \to e\pi^{\circ}N^{*}$:
    the proton is excited into an (I=1/2)-isobar while a high 
    energy single $\pi^{\circ}$ is produced by photon-Odderon fusion.
    \label{fig:feyn}}
\end{figure}

Since hadron-hadron scattering at high energies is generally dominated
by Pomeron exchange, an Odderon contribution is best searched for in
final states with quantum numbers to which Pomeron exchange cannot
contribute.  One possibility is the exclusive production of
pseudoscalar mesons at HERA via 
photon-Odderon 
fusion. The measurement
presented here uses the H1 detector~\cite{Abt:hi} to study exclusive
$\pi^{\circ}$ photoproduction in the reaction (see Fig.~\ref{fig:feyn})
\begin{equation}
  \label{eq:process}
  ep \to e\pi^{\circ}N^{*},
\end{equation}
where the photon virtuality is kept very small. The proton is
excited to an (I = 1/2)-isobar with negative parity, which
subsequently decays into a final state containing a highly energetic
neutron.  
In this exclusive reaction the scattered electron, the two photons from
the $\pi^{\circ}$ decay, and the leading neutron from the $N^{*}$
decay are detected. The remaining decay products of the $N^{*}$ go
undetected.

A calculation by Berger et al.~\cite{Berger:1999ca} predicts a sizeable
cross section for the
photoproduction process $\gamma p\to \pi^{\circ}N^{*}$.  
For this prediction a model in the framework of
non-perturbative QCD, the Stochastic Vacuum Model
(SVM)~\cite{Dosch:ha}, was extended and applied to high energy
scattering by functional methods~\cite{Nachtmann:ua}.  
The proton is treated as a quark-diquark system in transverse
space.  
A large variety
of high energy reactions has been described successfully with this
model, including data from HERA~\cite{Donnachie:2001wt}.  
For $\gamma p\to \pi^{\circ}N^{*}$,
a cross section of about 300~nb  is predicted \cite{Berger:1999ca} 
at a photon-proton centre of mass energy
of $W=20$~GeV, with an uncertainty of about a factor 
of 2 ~\cite{Berger:1999ca,Berger:2000wt}.
The energy dependence of the process is not predicted by the model. 
However, assuming that Odderon exchange leads to a cross section that is
flat or rises with energy, 
the cross section at HERA is expected to be at least 300~nb.
For an energy dependence $\propto (W^2)^{0.15}$ \cite{Berger:1999ca}
the
cross section at HERA would be a factor of approximately two larger than
that at $W = 20 \ {\rm GeV}$.

\section{Detector Description \label{chap:detdes}}
The analysis presented here is based on data taken with the H1
detector~\cite{Abt:hi} at HERA in 1999 and 2000 where electrons (or
positrons) with an energy of 27.5 GeV collided with protons of 920 GeV
energy. The data used for this analysis correspond to an integrated
luminosity of $30.6~\pbarnt$. In the following a short overview is
given of the essential detector components of H1 used in this
analysis.

Electrons are identified at $z = -33.4$~m in the electron
tagger~\footnote{The proton beam points to the ``forward'' ($+z$)
direction, where $z=0$ corresponds to the nominal interaction
point. Polar angles $\theta$ are measured with respect to this
direction.} which is a crystal Cherenkov calorimeter with 49 channels,
a total transverse size of 15.4 $\times$ 15.4 cm and a depth of 22
radiation lengths.

The study presented here is the first published analysis based on the
Very Low $\qsq$ calorimeter (``VLQ'')~\cite{Keller:1998at}.
Originally constructed for the detection of
scattered electrons in the transition region between the deep
inelastic scattering (DIS) and photoproduction regimes, the VLQ
is sensitive in the range $0.02<Q^{2}<1\;\gevsq$. Here $Q^{2}$ is the
modulus of the squared four-momentum transfer between the incoming and
scattered electrons. The VLQ is
used here as a photon detector.  
It is situated at $z = -3.02$~m and
covers the polar angular range $177.3\dgr < \theta < 179.4\dgr$.  The
VLQ is a tungsten-scintillator strip sandwich calorimeter with a
``projective readout''~\cite{Keller:1998at}. Its total thickness
amounts to 16.7 radiation lengths, and its Moli\`ere radius is
1.25~cm.  It consists of two identical modules which are located above
and below the beam pipe. Each module is read out at either end by
photodiodes.  The energy and position resolution for electromagnetic
showers are $\sigma_E/E=0.19/\sqrt{E/\gev} \oplus 0.064
\oplus 0.23/(E/\gev)$ and
$\sigma_x=\sigma_y=2.1\mathrm{mm}/\sqrt{E/\gev}$, respectively. The
double photon resolution is 1.5~cm, which is sufficient to separate
the photons from the decay of a 50~GeV $\pi^{\circ}$.  This distance
is much smaller than the 
minimal separation in the VLQ of 4~cm for the two 
photons from decays of $\pi^{\circ}$ mesons
with an actual maximum energy of 20~GeV.
From an investigation of samples of QED Compton events ($ep\rightarrow
ep\gamma$) the absolute positions of the VLQ modules are known to
better than 1~mm and the energy scale is determined with an uncertainty
of $\pm$ 4\%~\cite{Kluge}.

The SpaCal (``Spaghetti Calorimeter'')~\cite{spa} is a
lead-scintillating fibre calorimeter which is positioned at $z \approx
-1.55$~m and covers the polar angular range $153\dgr < \theta <
178\dgr$ with an energy resolution of $\sigma_E/E=0.075/\sqrt{E/\gev}
\oplus 0.010$, a polar angular resolution better than 2.5~mrad for
energies above 1 \gev and an energy scale uncertainty of $\pm$ 4\%.

The Forward Neutron Calorimeter (FNC)~\cite{Adloff:1999yg}, located at
$z=+107$ m in the HERA tunnel, detects high-energy neutrons.  The
acceptance, determined using inclusive events with a leading
neutron~\cite{Adloff:1999yg}, is $\approx$ 90\% for scattering angles
of $\theta \lesssim$ 0.1 mrad and vanishes above 0.6 mrad.

The tracking system consists of 2 m long coaxial cylindrical central
drift chambers ($25\dgr < \theta < 155\dgr $), a forward tracking
detector ($7\dgr < \theta < 25\dgr$) and a backward drift chamber in
front of the SpaCal.  The Liquid Argon (LAr) calorimeter ($4\dgr <
\theta < 154\dgr$) surrounds the central and the forward trackers.

\section{Event Selection \label{chap:evsel}}
The relevant Lorentz-invariant kinematical variables for
process~(\ref{eq:process}) are $Q^{2}$, the inelasticity $y$ and the
squared four-momentum transfer $t$ at the nucleon vertex.  The
quantity $y$ denotes the fractional energy transfer from the electron
to the proton in the rest frame of the proton and is calculated as
$y=(q\cdot p)/(k\cdot p)\approx 1-E'/E$ where $q$, $p$ and $k$ are the
four-momenta of the quasi-real photon, the target proton and the
incident lepton and $E$ ($E'$) is the energy of the incoming
(scattered) electron. The variable $t=(p-X)^{2}$, where $X$ is the
four-momentum of the outgoing $N^*$, can be reconstructed from the
squared transverse momentum of the $\pi^{\circ}$ candidate as
$t \simeq -h_{\perp}^2$ (see Fig. \ref{fig:feyn}).

Candidate events for the reaction~(\ref{eq:process}) are selected 
through the detection of an electron scattered through a very small 
angle, of two photons with combined invariant mass consistent 
with a $\pi^{\circ}$, and of a high energy neutron in the forward
direction. 
Scattered electrons with energies between 8.25 GeV and 19.25 GeV,
corresponding to $0.3<y<0.7$ and $Q^{2}<0.01\;\gevsq$, were selected
with the electron tagger. 
Odderon-induced $\pi^{\circ}$ production is expected very
close to the beam pipe in the backward direction due to the small
values of $t$ and the large photon energy. 
Two
electromagnetic calorimeters covering different regions in polar
angle in the backward region are therefore used to detect  
photons from the 
$\pi^{\circ}$ decay,
reconstructed as two separate clusters.
Photons in the SpaCal or the VLQ are selected by requiring a narrow
cluster with an energy well above the noise levels, i.e. larger than
90 MeV or 2 GeV, respectively. For trigger reasons at least one of the
photons must be reconstructed in the VLQ, with a total energy of at
least 6 \gev in one VLQ module. The intermediate excited nucleonic
state $N^{\ast}$ is selected by demanding a neutron in the FNC with an
energy above 200 \gev.

No activity 
is allowed
in the central detectors of H1, i.e. the tracking chambers
and the Liquid Argon calorimeter, and no additional energy deposition
apart from the two photon candidates is allowed
in the VLQ or SpaCal calorimeters. Events with charged
particles measured in the central tracking detectors in the range
$20\dgr <\theta < 160\dgr$ are rejected.  Due to the absence of
charged particles in the selected exclusive $\pi^{\circ}$ candidate
events, the interaction vertex cannot be reconstructed, and the event
kinematics are calculated using the mean interaction vertex as the
origin.

If no particles escape undetected, the variable
$\sum_{i}(E-P_{z})_{i}$, where $i$ runs over all final state particles
detected in the backward direction, namely the scattered electron and
the two photons from the $\pi^{\circ}$ decay, assumes a value equal to
twice the electron beam energy within detector resolution effects. A
cut of $49\;\gev<\sum_{i=e',\gamma ,\gamma}(E-P_{z})_{i}< 60\;\gev$
serves to reject events with additional particles emitted unobserved
in the backward direction, including photons from QED radiation.  For a
more detailed description of the event selection see~\cite{golling}.

\section{Monte Carlo Models}
The process (\ref{eq:process}) and its expected backgrounds are
simulated using the OPIUM~\cite{opi} and PYTHIA
\cite{Sjostrand:1993yb} event generators. The OPIUM generator is
derived from DIFFVM~\cite{dif}, which was originally designed to
simulate exclusive vector mesons produced by Pomeron exchange.  This
generator has been extended to OPIUM to include Odderon exchange with
an exclusive $\pi^{\circ}$ in the final state according to the
prescription in~\cite{Berger:1999ca}. The $t$ dependence of the cross
section is as given in~\cite{Berger:1999ca} and is approximately
proportional to $e^{bt}$ with a slope of $b=5.44\;\gev^{-2}$.

PYTHIA~\cite{Sjostrand:1993yb} is used to simulate the background from
inclusive $\gamma p$ interactions which mainly consists of
low multiplicity events with a
neutral pion in the final state together with further unobserved particles, 
for example events from exclusive $\omega$ or
$\rho^{\circ}$ photoproduction where the vector meson decays into
$\pi^{\circ}\gamma$. 
Contributions to elastic single $\pi^{\circ}$
production from Reggeon exchange
($\omega$-trajectory\footnote{Measurements~\cite{Braunschw} at low
energies ($W \approx 3$ GeV) were extrapolated to HERA energies}.) or
$\gamma\gamma$ fusion (``Primakoff effect'')~\cite{Berger:1999ca} are
negligible.
 
Since the hadronisation model applied in PYTHIA gives rise to a few
processes which do not 
conserve isospin, a modified version of the
program, referred to as ``PYTHIA-mod'', is also used for the
background description.  
Here, all processes violating isospin
conservation are excluded.  It is expected that the background is
bounded by the predictions of these two versions of the
model~\cite{sjo}. All Monte Carlo samples went through the same
reconstruction procedure as the data.

\section{Results}
\begin{figure}[htp]
  \begin{center}
  \epsfig{file=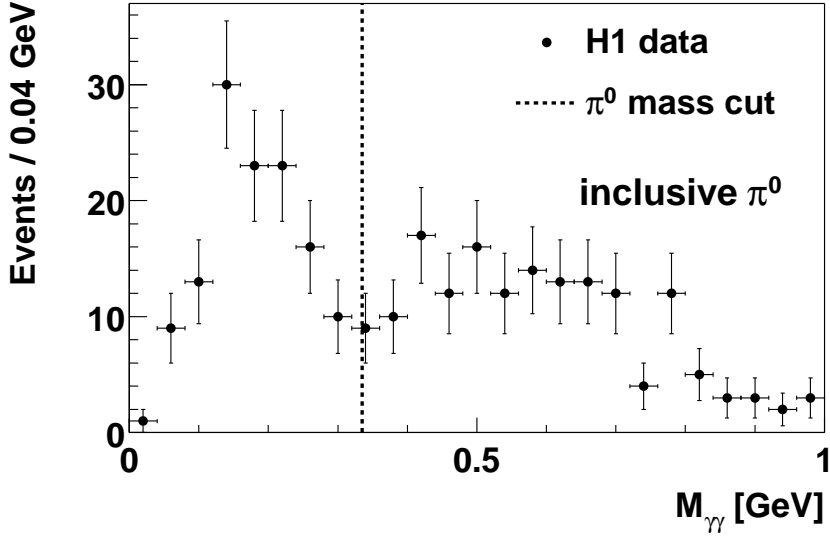,width=0.7\linewidth,clip=}
  \caption{Invariant mass distribution of two-photon candidates for
  all events with both photons in the VLQ,
  or one photon in the VLQ and one photon in the SpaCal. No restrictions
  are made on additional particles produced in the phase space region not
  covered by the VLQ and SpaCal (for full selection
  criteria, see text). The dashed line indicates the cut on $M_{\gamma \gamma}$
  applied in the exclusive analysis. \label{fig:incl}} 
\end{center}
\end{figure}

In order to demonstrate the capability to reconstruct $\pi^{\circ}$'s
in the backward calorimeters using the nominal interaction vertex
only, Fig.~\ref{fig:incl} shows the two-photon invariant mass
distribution for all events with two photons reconstructed in the
VLQ, or one photon in the VLQ and one in the SpaCal, as well as a
scattered electron in the electron tagger and a neutron in the
FNC. The additional veto cuts on the activity in the central detectors
of H1 and the variable $\sum_{i}(E-P_{z})_{i}$ were not applied for
this sample. A clear $\pi^{\circ}$ signal is observed. 
On the basis of this sample,
two-photon
candidates with combined invariant mass in the range
$M_{\gamma\gamma}<335$ MeV
are accepted as neutral pion candidates.

\begin{figure}[htp]
  \begin{center}
  \epsfig{file=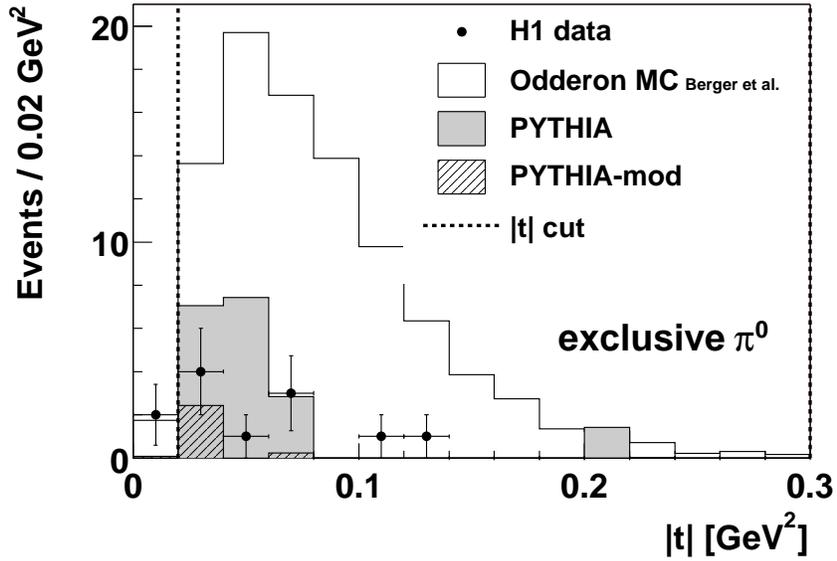,width=0.7\linewidth,clip=}
  \caption{Measured $t$ distribution for Odderon candidate events with
  $M_{\gamma\gamma}<$~335 MeV. The background expectation from PYTHIA
  and PYTHIA-mod are also shown together with the predicted number of
  Odderon-induced events~\cite{Berger:1999ca}. In the final selection
  the two events in the first bin are rejected due to the acceptance cut
  $|t|>0.02\;\gevsq$. \label{fig:t}} \end{center}
\end{figure}

Figure~\ref{fig:t} shows the $|t|$ distribution for exclusive
$\pi^{\circ}$ candidate events with $M_{\gamma\gamma}<335$ MeV after the 
full event selection. 
Mainly due to the limited angular coverage of the VLQ,
the acceptance in $|t|$ vanishes at very small $|t|$,
reaches a maximum at $|t|\approx 0.05$ \gevsq and drops again at
larger $|t|$ values.
A cut of $0.02 <|t|<0.3\;\gevsq$ is used to
define the accessible $|t|$ range and changes the predicted cross
section \cite{Berger:1999ca}
$\sigma(\gamma p\to\pi^{\circ}N^{\ast})$ via $\gamma\odd$
fusion 
by a factor of approximately 2/3, such that the measurable cross section
is expected to remain above 200~nb.

\begin{figure}[htp]
  \begin{center}
  \epsfig{file=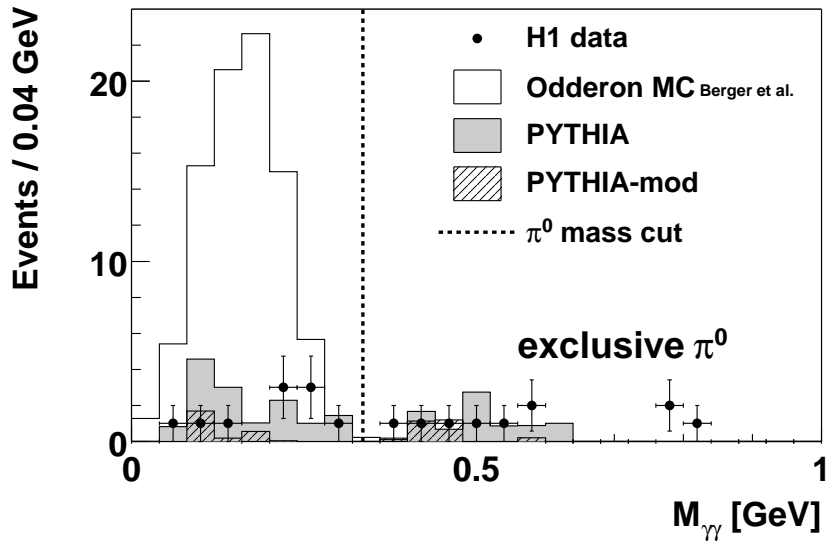,width=0.7\linewidth,clip=}
  \caption{Invariant mass distribution of two-photon candidates for
  exclusive events with both photons in the VLQ,
  or one photon in the VLQ and one photon in the SpaCal (for selection
  criteria see text). The backgrounds computed from the PYTHIA and
  PYTHIA-mod models are also shown together with the distribution for
  Odderon exchange predicted from~\cite{Berger:1999ca} where the expe\-ri\-mentally observed
  width of the $\pi^{\circ}$ signal is taken from the inclusive
  $\pi^{\circ}$ sample (see Fig.~\ref{fig:incl}).  \label{fig:mass}} \end{center}
\end{figure}

Figure~\ref{fig:mass} shows the
distribution of the two-photon invariant mass $M_{\gamma\gamma}$ after
the complete event selection, including the $|t|$ cut. A total of 10
events containing two-photon candidates with invariant mass
$M_{\gamma\gamma}<335$ MeV remains, and no $\pi^{\circ}$ peak is
observed. 
The background estimates from PYTHIA and PYTHIA-mod are 12 and 3
events, respectively.  
The measured data are found to be
consistent with the simulated background from PYTHIA or PYTHIA-mod,
both in magnitude and shape. 
By varying the
normalization of the PYTHIA-mod simulation by $\pm$ 100\% and by
estimating the background from the PYTHIA-mod prediction or the data
for $M_{\gamma\gamma} >$ 335~MeV, the background is estimated to be
below 12 events. 
From the photon-Odderon fusion
model~\cite{Berger:1999ca} described above, assuming the cross section
has no dependence on $W$, 90 events are expected. 

In order to reduce the model dependence
in limit calculations, the PYTHIA predictions are
disregarded and a background of zero events is assumed.  Within this
most conservative scenario an upper limit for the cross section of
reaction~(\ref{eq:process}) is determined, using the statistical
method described in~\cite{Feldman:1997qc,Cousins:1991qz}.  In addition
to the statistical uncertainties in the data a systematic uncertainty
of 25\% is taken into account. The latter arises mainly from the
acceptances of the forward neutron calorimeter (20\%), the electron
tagger (5\%) and the VLQ (4\%).  To calculate the limit for negative
parity $N^*$ production the four dominant states are considered:
$N$(1535) and $N$(1650) (pion and neutron in relative $S$-wave), and
$N$(1520) and $N$(1700) (pion and neutron in $D$-wave), as also used
in the theoretical estimate~\cite{Berger:1999ca}. Higher mass
negative-parity states and nucleon resonances with positive parity
might also contribute to the final event sample.
Since these
contributions are not subtracted, 
the limit derived from the sample
is a conservative upper bound to be compared to the theoretical
prediction. Table~\ref{tab:acceff} summarizes the branching ratios,
acceptances and efficiencies used to determine the limit on the
$\gamma p$ cross section, which is
extracted~\cite{vonWeizsacker:sx,Frixione:1993yw} by calculating
a limit for the
$ep$ cross section and dividing by a photon flux factor integrated
over the kinematic region $Q^2 < 0.01$ GeV$^2$ and $ 0.3 < y < 0.7$.
Assuming a slope of $b=5.44$~GeV$^{-2}$ for the differential cross
section $d\sigma/dt$ as given in~\cite{Berger:1999ca}, the limit for
the photoproduction cross section integrated over the accessible $|t|$
range $0.02<|t|<0.3\;\gevsq$ is
\begin{equation}
  \label{eq:limit}
  \sigma_{\gamma p \to \pi^{\circ}N^{*}}(\gamma\odd\;\mathrm{fusion}) <
  49\; \mathrm{nb \hspace{1.5cm} (95~\%~CL)}
\end{equation}
at an average $\gamma p$ centre-of-mass energy $\langle W \rangle =$
215 GeV.  The limit changes by $+$29\% ($-$17\%) for a slope $b$ of
3~GeV$^{-2}$ (8 GeV$^{-2}$) instead of 5.44 GeV$^{-2}$.

The derived limit is clearly incompatible with the 
predicted value ~\cite{Berger:1999ca} of
at least
200~nb for this kinema\-ti\-cal range at HERA energies.

\section{Conclusion and Outlook}
A first search for events produced through Odderon exchange in the
process $ep \to e\pi^{\circ}N^{*}$ is reported in the kinematical
range 174 $<W<$ 266 GeV, $Q^{2}<0.01\;\gevsq$ and
$0.02<|t|<0.3~\;\gevsq$.  
No
$\pi^{\circ}$ signal is observed and 
the number of reconstructed events
containing two photons with invariant mass in the $\pi^{\circ}$ region
is found to be compatible with background estimates. 
An upper limit for the photoproduction
cross section of $ \sigma_{\gamma p \to
\pi^{\circ}N^{*}}(\gamma\odd\;\mathrm{fusion}) < 49\;
\mathrm{nb} $ is derived, which depends only weakly on the details of
the production mechanism.

An Odderon-induced process for exclusive $\pi^{\circ}$ production of
the magnitude predicted by Berger et al.~\cite{Berger:1999ca} is not
compatible with the data.  There are two possible interpretations of
this result within the assumptions of~\cite{Berger:1999ca}. The first
is that the Odderon intercept $\alpha_{\oddsmall}(0)$, which
characterizes the energy dependence of the cross section, is
considerably smaller than that of the Pomeron.  A value of
$\alpha_{\oddsmall}(0)<0.7$ would be compatible with the measurement
and with alternative predictions~\cite{Kaidalov:1999de}.  The second
interpretation is that the process is of diffractive nature but that
the coupling at the $\gamma\odd\pi$-vertex is smaller than anticipated
in~\cite{Berger:1999ca}.  Further insight might come from a search for
the production of heavier tensor mesons \cite{Berger:2000wt}
or from charge asymmetry
measurements in exclusive $\pi^+ \pi^-$ production \cite{Ivanov:2001zc}
or charm production~\cite{Brodsky:1999mz}.

\section*{Acknowledgements}

We are grateful to the HERA machine group whose outstanding efforts
have made and continue to make this experiment possible. We thank the
engineers and technicians for their work in constructing and
maintaining the H1 detector, our funding agencies for financial
support, the DESY technical staff for continual assistance and the
DESY directorate for the hospitality which they extend to the non-DESY
members of the collaboration. In particular the authors wish to thank
H.~G.~Dosch, O.~Nachtmann and T.~Sj\"ostrand for stimulating
discussions.


\vspace*{1.5cm}

\begin{table}[htp]
  \begin{center}
    \begin{tabular}{|l|c|}\hline
      BR$(N^{*}\to n+X)$ & (42 $\pm$ 3) \% \\ \hline
      BR$(\pi^{\circ}\to\gamma\gamma)$  & (98.80 $\pm$ 0.03) \% \\ \hline
      neutron acceptance  & (11 $\pm$ 2) \% \\ \hline
      $\pi^{\circ}$ acceptance     & (6.0  $\pm$ 0.3) \% \\ \hline
      electron acceptance    & (40 $\pm$ 3) \%  \\ \hline
      FNC trigger efficiency & (98 $\pm$ 1) \% \\ \hline
      VLQ trigger efficiency & (95 $\pm$ 2) \% \\ \hline
      photon flux factor  & 0.0136 \\ \hline
\end{tabular}
\caption{Summary of branching ratios, acceptances and efficiencies 
with the respective errors necessary for the determination of the 
limit on the cross section~(\ref{eq:limit}).
    \label{tab:acceff}}
\end{center}
\end{table}

\end{document}